\documentclass[aps,pre,reprint,superscriptaddress]{revtex4-1}
\usepackage{graphicx}
\usepackage{dcolumn}
\usepackage{subfigure}

\begin{document}

\title{Intrinsic 2D ferromagnetism, quantum anomalous Hall conductivity, and fully-spin-polarized edge states of FeBr$_3$ monolayer}

\author{Shi-Hao Zhang}
\affiliation{Beijing National Laboratory for Condensed Matter Physics, Institute of Physics, Chinese Academy of Sciences, Beijing 100190, China}
\affiliation{School of Physical Sciences, University of Chinese Academy of Sciences, Beijing 100190, China}
\author{Bang-Gui Liu}
\email[]{bgliu@iphy.ac.cn}
\affiliation{Beijing National Laboratory for Condensed Matter Physics, Institute of Physics, Chinese Academy of Sciences, Beijing 100190, China}
\affiliation{School of Physical Sciences, University of Chinese Academy of Sciences, Beijing 100190, China}

\date{\today}

\begin{abstract}
It is of great interest to explore intrinsic two-dimensional ferromagnetism and seek better two-dimensional quantum anomalous Hall insulator materials. Here, we show that the FeBr$_3$ monolayer is an intrinsic two-dimensional ferromagnetic material whose Curie temperature is 140 K thanks to its strong spin exchange interaction and giant uniaxial magnetic anisotropy. Our phonon spectra and mechanical analysis indicate that the FeBr$_3$ monolayer is dynamically and mechanically stable. Our electronic structure calculation shows that there is one Dirac cone at K point in the Brillouin zone and the spin-orbit coupling opens a semiconductor gap of 33.5 meV. Further tight-binding analysis reveals that the Chern number is equivalent to 1 and there is a quantum anomalous Hall conductivity $\sigma _{xy} = e^2/h$,  and the chiral edge states are fully spin-polarized when an edge is created. Furthermore, it is shown that the main results are not affected by electron correlation effects and biaxial strain. Therefore, this FeBr$_3$ monolayer as 2D material would be useful for spintronic applications.
\end{abstract}

\pacs{}

\maketitle


\section{Introduction}

Recent experimental discovery of intrinsic two-dimensional ferromagnetism~\cite{add1,add2,huang2017layer} attracts people's attention on atomic monolayers as two-dimensional materials. As the Mermin-Wagner theorem implies~\cite{PhysRevLett.17.1133}, it is very difficult to truly realize two-dimensional ferromagnetism at non-zero temperature. In addition to strong spin exchange interactions, a strong uniaxial magnetic anisotropy is necessary to achieve a high Curie temperature. Most importantly, key properties must be robust against strain in such two-dimensional materials, to make heterostructures for practical devices. If the spin-orbit coupling opens a global gap and there exists non-zero Chern number indicating nontrivial topological properties, quantum anomalous Hall effect can be achieved in such two-dimensional ferromagnetic (or ferrimagnetic) materials. It will be of much interest to achieve stable intrinsic 2D ferromagnetism with high Curie temperature and high spin polarization.

Quantum anomalous Hall (QAH) effect was first theoretically predicted by introducing magnetism into two-dimensional honeycomb model~\cite{PhysRevLett.61.2015}, where the magnetism breaks the time reversal symmetry and external magnetic field is not necessary. Then, transition metal (TM) doped topological insulators~\cite{yu2010quantized}, graphene-based materials ~\cite{PhysRevLett.108.056802,PhysRevLett.112.116404,PhysRevB.92.165418,PhysRevLett.113.256401}, quantum wells~\cite{PhysRevLett.112.096804}, and heterostructures~\cite{TiO2,PhysRevLett.110.116802} are theoretically predicted to achieve the QAH effect. In these systems, the transition metal atoms can make the magnetism and the spin-orbit coupling (SOC) can open the global band gap, and then appear QAH conductivity and dissipationless chiral edge states. These features can help the QAH insulators to find applications in the low-power-consumption electronic devices~\cite{QAH,QAH2},
but on the experimental side, the QAH effect has only been observed in Cr or V doped (Bi,Sb)$_2$Te$_3$ thin film at very low temperature ($<$85 mK)~\cite{chang2013experimental,PhysRevLett.113.137201,chang2015high}. It has been theoretically predicted that some transition-metal compound monolayers can host the quantum anomalous Hall effect~\cite{NiRuCl6,C6NR08522A,PhysRevB.95.045113,PhysRevB.95.201402,cp},
but it is still highly desirable to seek better, experimentally realizable two-dimensional QAH insulator materials, especially with fully-spin-polarized edge states.

Here, we propose that FeBr$_3$ monolayer can host intrinsic 2D ferromagnetism, quantum anomalous Hall effect, and fully-spin-polarized edge states. Our investigation shows that it is dynamically and mechanically stable, and its Curie temperature is 140 K thanks to the strong perpendicular uniaxial magnetic anisotropy. There is one Dirac point at the K point, and the spin-orbit coupling creates a global gap of 33.5 meV. Our tight-binding calculation with maximally localized Wannier functions reveals that the quantum anomalous Hall conductivity is $\sigma_{xy}=e^2/h$ and there are fully-spin-polarized chiral edge states when a edge is made. The key properties of the system remain robust when biaxial strain is applied. The more detailed results will be presented in the following.

\section{Computational methods}

First-principles spin-polarized calculations are done with the projector-augmented wave (PAW) method~\cite{PhysRevB.50.17953} as implemented in the Vienna Ab initio Simulation Package (VASP)~\cite{PhysRevB.47.558}. The generalized gradient approximation (GGA) by Perdew, Burke, and Ernzerhof~\cite{PhysRevLett.77.3865} is used for the exchange-correlation potential. The Brillouin zone integration is carried out with a $\Gamma$-centered (15$\times$15$\times$1) Monkhorst-Pack grid~\cite{PhysRevB.13.5188}. The structures are fully optimized to ensure that all the Hellmann-Feynman forces on each atom are less than 0.01 eV/\AA{} and the total energy difference between two successive steps is smaller than $10^{-6}$ eV. To ensure the structural stability of the monolayers, phonon spectra are calculated in terms of the density functional perturbation theory as implemented in the PHONOPY program~\cite{TOGO20151}. The spin-orbit coupling is taken into account to study the magnetocrystalline anisotropy. Band dispersion calculations with GGA+U~\cite{PhysRevB.57.1505} functional are carried out to make further confirmation. The tight-binding Hamiltonian is constructed with the help of maximally localized Wannier functions (MLWFs)~\cite{RevModPhys.84.1419} from the DFT calculated bands, as implemented in the Wannier90 code~\cite{MOSTOFI2008685}. The surface state spectrum of a semi-infinite system is obtained by the surface Green's function~\cite{GF1,GF2}. The calculation of Berry curvature is carried out with a denser k mesh (121$\times$121$\times$1).

\section{results and discussion}

\subsection{Structure and stability}

\begin{figure}[!htbp]
\includegraphics[width=0.48\textwidth]{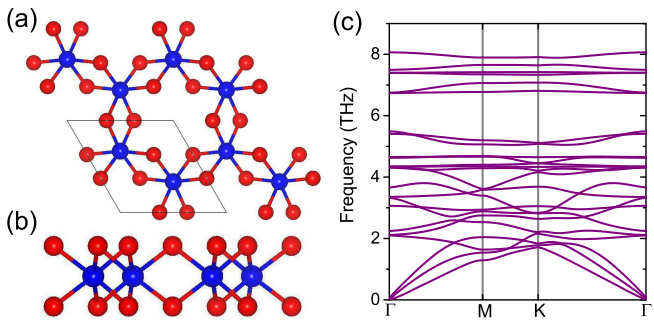}
\caption{~\label{fig1} The top view (a), side view (b), and phonon spectra (c) of the FeBr$_3$ monolayer. The red and blue balls represent Br and Fe atoms, respectively.}
\end{figure}

On experimental side, bulk FeBr$_3$ assumes a layered structure like graphite, and each of FeBr$_3$ monolayers is similar to graphene and the monolayers are bound through van der Waals interaction. As for magnetic ordering, it is ferromagnetic within the monolayer and antiferromagnetic between the nearest monolayers. We study one FeBr$_3$ monolayer in terms of first-principles investigation. The calculational model is constructed by repeating the FeBr$_3$ monolayer and adding a vacuum layer between adjacent FeBr$_3$ monolayers. The optimized structure of FeBr$_3$ monolayer is like that of CrI$_3$ monolayer. Each Fe is coordinated by six Br atoms with Fe-Br bond length of 2.42 \AA{}, and the Fe atomic layer is sandwiched between the two Br layers (the Br-Br plane distance is the 2.75 \AA{}), as shown in Fig. 1. The unit cell includes two Fe atoms and six Br atoms, and the equilibrium lattice constant is 6.29 \AA{}. The Fe atoms form a honeycomb lattice, with the Fe-Fe distance being 3.63\AA, and the two nearest Fe atoms share two Br atoms belonging to different Br layers. The Fe d electrons make a local magnetic moment of 1$\mu _B$, staying in the low-spin state, and the ground-state phase is ferromagnetic because the antiferromagnetic structures are higher by at least 75 meV per formula unit.

The formation energy, defined as $E_f=E_{\rm FeBr_3}-1/4\mu _{\rm Fe}-3/4\mu _{\rm Br}$, is calculated, being equivalent to -0.30 eV/atom. The negative value favors experimental synthesis. We also calculate the interlayer binding energy by $E_b = (E_{\rm monolayer}-E_{\rm bulk}/N)/S$, where $E_{\rm monolayer}$ and $E_{\rm bulk}$ are the total energies of the monolayer and bulk FeBr$_3$ per unit cell, $S$ is the area of monolayer per unit cell, and $N$ is the number of layers in bulk FeBr$_3$. Our calculated binding energy of FeBr$_3$ monolayer is 11.3 meV/\AA{}$^2$, smaller than experimental value of graphene (23.3 meV/\AA{}$^2$, vs. graphite)~\cite{PhysRevB.69.155406}. This implies that the FeBr$_3$ monolayer can be exfoliated from bulk FeBr$_3$ material. In Fig. 1c, we also present the phonon spectra of the FeBr$_3$ monolayer. It is clear that there exists no imaginary frequencies, which ensures the dynamical stability of the FeBr$_3$ monolayer.

To evaluate the mechanical stability of the FeBr$_3$ monolayer, we calculate the elastic moduli, obtaining $C_{11} = 35.9$ N/m, $C_{22} = 36.1 $ N/m, $C_{12} = 11.6$ N/m, and $C_{66} = 12.9$ N/m. The Young's module, nearly isotropic in the plane, is equivalent to 32 N/m. They satisfy the criteria of stability of two-dimensional materials ($C_{11}C_{22} > C_{12}C_{21}$ and $C_{66} > 0$)~\cite{PhysRevB.85.125428}. For the square flake, the ratio between the out-of plane deformation $h$ induced by its own gravity and the edge length $l$ is $h/l \approx (\rho gl/Y)^{1/3}$, where $g$ being the gravitational acceleration and $\rho$ the density of two-dimensional materials (2.89$\times$10$^{-6}$ kg/m$^2$). When $l$ is 100 $\mu$m, the ratio becomes 4.4$\times$10$^{-4}$ which is so small that the FeBr$_3$ monolayer can keep its stability without the support of a substrate.

\subsection{Electronic structure}

The spin-resolved density of states (DOS) of the FeBr$_3$ monolayer are presented in Fig. 2. The triangular antiprismatic crystal field  splits Fe atom's d orbitals into a triplet (d$_{xy}$, d$_{x^2-y^2}$, and d$_{z^2}$) and a doublet (d$_{yz}$ and d$_{xz}$). For the spin-up channel, the triplet DOS is peaked at -0.7 eV and the doublet DOS at 0.8 eV; and for the spin-down channel, they are peaked at -0.3 and 1.2 eV, respectively. In terms of the crystal field theory, the relative strength of the crystal field splitting ($\Delta E_{cf}\sim 1.5$ eV) and spin exchange splitting ($\Delta E_{ex}\sim 0.4$ eV) leads to a low spin state for Fe atom. Each Fe atom has five valence electrons  and they all occupy the triplet bands: three for spin-up and two for spin-down. These makes $1\mu _{B}$ for a Fe atom.
The energy bands  without the spin-orbit coupling (SOC) are presented in FIG. 3a. The spin-up energy bands show the feature of semiconductor with direct gap of 1.23 eV, and the valence band maximum (VBM) and conduction band minimum (CBM), both located at the M point, are well separated from the Fermi level. As for the spin-down energy bands, it is clear that the six triplet bands (for two Fe atoms) are located in the spin-up gap, and there is one Dirac point (mainly from d$_{xy}$ and d$_{x^2-y^2}$ orbitals) located at the K point. Therefore, the FeBr$_3$ monolayer is a half metal, implying that it is fully spin polarized. Our further test calculations show that this electronic structure is robust when a correlation $U$ parameter is introduced and varies from 0.5 eV to 2 eV.

\begin{figure}[!htbp]
\includegraphics[width=0.45\textwidth]{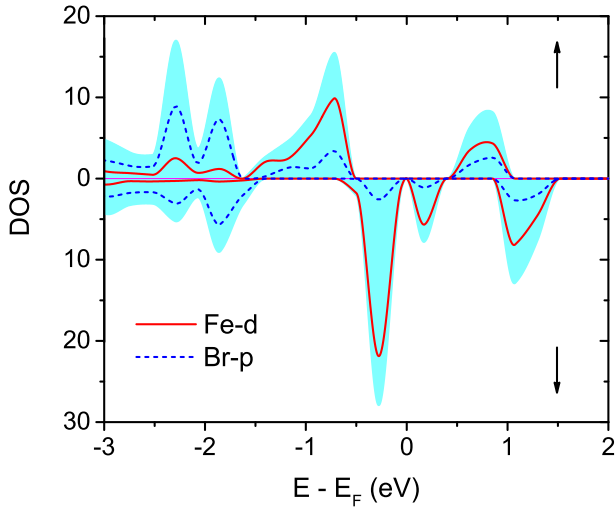}
\caption{~\label{fig3} The total (shadow) and partial (solid and dash lines) density of states (DOS) of the FeBr$_3$ monolayer. }
\end{figure}

\begin{figure}[!htbp]
\includegraphics[width=0.4\textwidth]{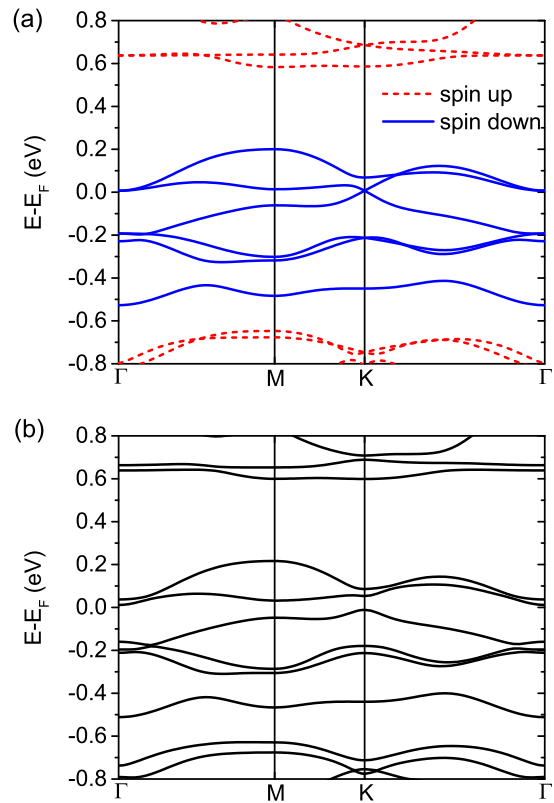}
\caption{~\label{fig2} The calculated energy bands of the FeBr$_3$ monolayer without (a) and with (b) the spin-orbit coupling. For the spin-up channel, there is a gap of 1.2 eV near the Fermi level.}
\end{figure}

The band structure with SOC taken into account is presented in Fig. 3b. It is important that there is a semiconductor gap open at the Fermi level and therefore the FeBr$_3$ monolayer is a barrow-gap semiconductor.
The gap 65 meV at the Dirac point can be attributed to the large SOC effect from Fe d orbitals under the special octahedral environment. The energy degeneracy at the $\Gamma$ point near the Fermi level is also lifted, which makes the lower-energy band at the $\Gamma$ point become the new conduction band minimum. As a result, the global semiconductor gap of the FeBr$_3$ monolayer is 33.5 meV. We also confirm the key results by considering the $U$ parameter. When $U$ reaches to 2 eV, the FeBr$_3$ monolayer is still a semiconductor with a global gap of 60 meV.

\subsection{Intrinsic 2D ferromagnetism}

The ferromagnetic order comes from the Fe moments of 1$\mu _B$ and the Fe spins form a honeycomb lattice of 3.63\AA. According to the Goodenough-Kanamori-Anderson (GKA) rules~\cite{PhysRev.100.564,Kanamori,PhysRev.115.2}, the superexchange interaction from d-p-d path usually tends to be antiferromagnetic if the cation-anion-cation bond angle is 180$^{\circ}$, but it can be ferromagnetic if the cation-anion-cation bond angle is near 90$^{\circ}$. The FeBr$_3$ monolayer belongs to the latter because the Fe-Br-Fe angle is 97$^{\circ}$ which is close to 90$^{\circ}$. As a result, the inter-spin interaction in the FeBr$_3$ monolayer is ferromagnetic, even under biaxial strain up to 5\%.

If the 2D spin system is isotropic, Mermin-Wagner theorem~\cite{PhysRevLett.17.1133} implies that the strong thermal fluctuations of gapless long-wavelength modes will destroy the 2D ferromagnetism at finite temperature. In order to achieve true 2D magnetism at finite temperature, some uniaxial magnetic anisotropy is necessary. Our first-principles calculations reveal that the total energy of the FeBr$_3$ monolayer is dependent on the spin orientation. The out-of-plane direction is favorable and the in-plane direction is higher by 0.6 meV per formula unit, which is comparable to that of CrXTe$_3$ (X = Si,Ge,Sn, the magnetic anisotropic energies are 0.069-0.419 meV/f.u.)~\cite{PhysRevB.92.035407}. The magnetic anisotropic energy is originated from the SOC effect. Our first-principles calculation reveals that the z-component orbital moment of each Fe atom is $\langle L_z \rangle = 0.215 \hbar$. This large orbital moment can explain the large magnetic anisotropy.

The magnetic moment of 1$\mu _B$ means that the Fe spin is $s=1/2$ and $(S_{i}^{z})^{2}$ is always $1/4$, implying no single-ion anisotropy. Because of the crystal symmetry of the FeBr$_3$ monolayer, the effective spin Hamiltonian can be written as
\begin{equation}
H = -\sum_{ij}(J_{ij}\textbf{S}_i\cdot \textbf{S}_j+\lambda_{ij}S^z_iS^z_j)
\end{equation}
where the $J$ term is the isotropic spin exchange interaction and the $\lambda$ term describes the Ising exchange  anisotropy. It is expected that the exchange constant $J_{ij}$ and the Ising anisotropy $\lambda_{ij}$ are both limited to the nearest $\langle ij\rangle$ and next-nearest $\langle\langle ij\rangle\rangle$ neighbors, namely $J_{\langle ij\rangle}=J_1$, $\lambda_{\langle ij\rangle}=\lambda_1$, and $J_{\langle\langle ij\rangle\rangle}=J_2$, and $\lambda_{\langle\langle ij\rangle\rangle}=\lambda_2$. Actually, by comparing the total energies of ferromagnetic and anti-ferromagnetic states with the spins orientating parallel and perpendicular to the Fe plane, we obtain that $J_1$ = 48.4 meV, $J_2$ = -0.77 meV, and $\lambda _1$ = 1.70 meV, $\lambda _2$ = -0.45 meV. This is a two-dimensional quantum spin model with Ising anisotropy. It is not easy to obtain its accurate phase transition temperature, but we can obtain a lower limit to it by performing Monte Carlo simulation of the spin model with the classical spin approximation. We take the 100$\times$100 spin supercell to perform Monte Carlo simulation and thereby show that the Curie temperature $T_c$ is about 140 K. The spin wave gap $\Delta$ is 2.4 meV ($3\lambda _1+6\lambda_2$)~\cite{CrI3}, which guarantees the existence of 2D long-rang magnetic ordering at non-zero temperature.

\begin{figure}[!htbp]
\includegraphics[width=0.45\textwidth]{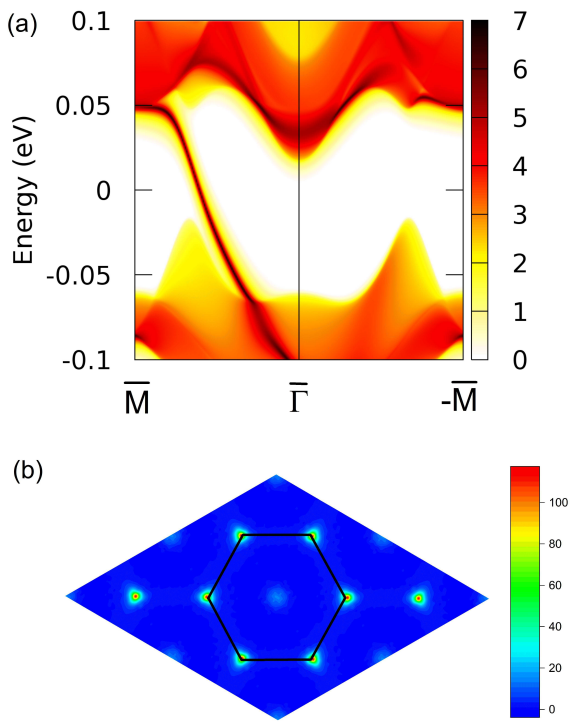}
\caption{~\label{fig4} (a) Tight-binding band structure of the semi-infinite sheet of the FeBr$_3$ monolayer, including the  edge states in the gap of the bulk bands; (b) The k-space distribution of the Berry curvature $\Omega_z(k)$ from all the occupied states, with the black hexagon denoting the first Brillouin zone.}
\end{figure}

\subsection{Quantum anomalous Hall effect}

To identify the topological properties of the FeBr$_3$ monolayer, we calculate the Berry curvature ($\Omega_z$) and Chern invariant ($C$) through the effective tight-binding model of the first-principles energy bands. The Berry curvature is defined by~\cite{PhysRevLett.95.156601,PhysRevLett.94.226601}
\begin{eqnarray}
\Omega_z(k)=-2\sum_{m \neq n}\frac{Im \langle \Psi _{nk}|v_x | \Psi _{mk}\rangle \langle \Psi _{mk}|v_y | \Psi _{nk}\rangle}{(E_{nk}-E_{mk})^2}
\end{eqnarray}
where $n$ and $m$ are the band indexes and the summation is done for the occupied bands. The Chern number is defined as the integral of the Berry curvature ($\Omega_z$) over the first Brillouin zone (BZ). Our calculation shows that the Chern number is 1. This means that the quantum anomalous Hall conductivity can be expressed as
\begin{equation}
\sigma_{xy} = \frac{e^2}{h}.
\end{equation}
Therefore, the FeBr$_3$ monolayer is a Chern insulator and can host quantum anomalous Hall effect.

Furthermore, we can construct a semi-infinite sheet of the FeBr$_3$ monolayer and investigate its edge states using the effective TB model. We present the edge states (and the bulk states) in Fig. 4a. It is clear that the edge states form a chiral band crossing the Fermi level, which is because the intrinsic ferromagnetism has broken the time reversal symmetry. It is very important that the chiral edge states are fully-spin-polarized, which is promising for spintronics applications.
We also show the k-resolved Berry curvature in FIG. 4b. It is interesting that the Berry curvature is mainly located at the K and $\Gamma$ points in the Brillouin zone.

\begin{figure}[!htbp]
\includegraphics[width=0.4\textwidth]{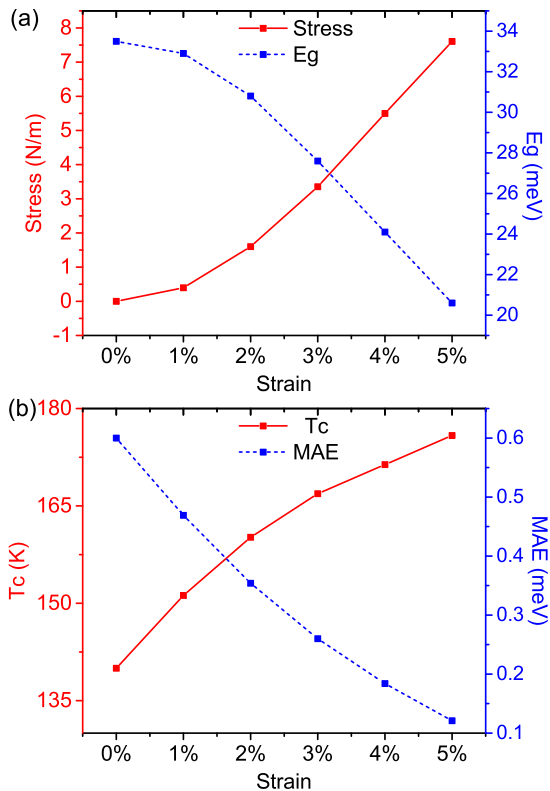}
\caption{~\label{fig5} The stress and global gap $E_{g}$ (a) and the Curie temperature $T_c$ and magnetic anisotropic energy (MAE) per formula unit (b) of the FeBr$_3$ monolayer as functions of the biaxial strain.}
\end{figure}

\subsection{Effect of biaxial strain}

We can investigate the robustness of topological features by applying biaxial strain to the FeBr$_3$ monolayer. With the strain $\eta$ given, the stress can be calculated by $\sigma = \partial E/ 2S\partial \eta$, where $S$ is the area of unit cell. The strain will change the SOC-induced semiconductor gap $E_g$, the Curie temperature $T_c$, and the magnetic anisotropy (MAE). We present the strain dependences of the stress, $E_g$, $T_c$, and MAE in Fig. 5. The semiconductor gap $E_g$ decreases with the strain, reducing to 20.7 meV at the strain of 5\%. The biaxial strain of 5\% corresponds to the stress of 7.6 N/m. When the strain is less than 5\%, we still have the Chern number, the quantum anomalous Hall effect, and the fully-spin-polarized edge states in the FeBr$_3$ monolayer. In addition, the biaxial strain enhances the Curie temperature $T_c$, but reduces magnetic anisotropic energy. It is interesting that the biaxial strain of 5\% effectively promotes the Curie temperature to 175 K.

\section{Conclusion}

In summary, we have shown that the FeBr$_3$ monolayer is an intrinsic two-dimensional ferromagnetic material whose Curie temperature is 140 K because the distorted octahedral crystal field leads to unquenched Fe orbital moment and then produces giant uniaxial magnetic anisotropy. Our phonon spectra and mechanical analysis indicates that the FeBr$_3$ monolayer is dynamically and mechanically stable. Our electronic structure calculation shows that there is one Dirac cone at K point in the Brillouin zone and the spin-orbit coupling opens a global gap of 33.5 meV. Further TB analysis reveals that the Chern number is equivalent to 1 and there is a quantum anomalous Hall conductivity $\sigma _{xy} = e^2/h$,  and there are fully-spin-polarized edge states when an edge is created. It has been shown that the main results are not affected by the Hubbard $U$ parameter, and also remain robust when a biaxial strain (up to 5\%) is applied. We believe that the FeBr$_3$ monolayer will be useful for spintronic applications.

\begin{acknowledgments}
This work is supported by the Nature Science Foundation of China (No.11574366), by the Strategic Priority Research Program of the Chinese Academy of Sciences (Grant No.XDB07000000), and by the Department of Science and Technology of China (Grant No.2016YFA0300701). The calculations were performed in the Milky Way \#2 supercomputer system at the National Supercomputer Center of Guangzhou, Guangzhou, China.
\end{acknowledgments}

\end{document}